\documentclass[11pt]{article}
\usepackage{amsfonts}
\usepackage{amsfonts}
\usepackage{amsfonts,amsmath,amssymb,epsf,indentfirst}
\usepackage{graphicx,color}
\usepackage[usenames,dvipsnames]{pstricks}
\usepackage{epsfig}
\usepackage{pst-grad} 
\usepackage{pst-plot} 
\usepackage{verbatim}
\numberwithin{equation}{section}

\newcommand{\al}{\alpha'}
\newcommand{\de}{\partial}
\newcommand{\be}{\begin{equation}}
\newcommand{\ba}{\begin{eqnarray}}
\newcommand{\ea}{\end{eqnarray}}
\newcommand{\ee}{\end{equation}}

\newcommand{\we}{\wedge}

\newcommand{\f}{\frac}
\newcommand{\s}{\sqrt}
\newcommand{\vp}{\varphi}

\newcommand{\ti}{\tilde}
\newcommand{\ap}{\alpha}

\newcommand{\ddd}{\cdot\cdot\cdot}
\newcommand{\no}{\nonumber \\}

\newcommand{\ep}{\epsilon}

\begin{document}

\begin{titlepage}
\thispagestyle{empty}
\renewcommand{\thefootnote}{\fnsymbol{footnote}}

\begin{flushright}
KEK-TH-1307 \\
IPMU09-0032
\end{flushright}

\bigskip

\begin{center}
\noindent{\Large \textbf{ABJM with Flavors and FQHE}}\\
\vspace{15mm} Yasuaki Hikida$^{a}$\footnote{e-mail:
hikida@post.kek.jp},\ Wei Li$^{b}$\footnote{e-mail:
wei.li@ipmu.jp}, and
Tadashi Takayanagi$^{b}$\footnote{e-mail: tadashi.takayanagi@ipmu.jp}\\
\vspace{1cm}

{\it $^{a}$High Energy Accelerator Research Organization (KEK), \\
 Tsukuba, Ibaraki 305-0801, Japan\\
 $^{b}$Institute for Physics and Mathematics of the Universe (IPMU), \\
 University of Tokyo, Kashiwa, Chiba 277-8582, Japan}

\vskip 3em
\end{center}

\begin{abstract}

We add fundamental matters to the ${\cal N}=6$ Chern-Simons theory
(ABJM theory), and show that D6-branes wrapped over $AdS_4 \times
S^3/\mathbb{Z}_2$ in type IIA superstring theory on $AdS_4\times
\mathbb{CP}^3$ give its dual description with ${\cal N}=3$
supersymmetry. We confirm this by the arguments based on R-symmetry,
supersymmetry, and brane configuration of ABJM theory. We also
analyze the fluctuations of the D6-brane and compute the conformal
dimensions of dual operators. In the presence of fractional branes,
the ABJM theory can model the fractional quantum Hall effect (FQHE),
with RR-fields regarded as the external electric-magnetic field. We
show that an addition of the flavor D6-brane describes a class of
fractional quantum Hall plateau transitions.

\end{abstract}

\setcounter{footnote}{0}

\renewcommand{\thefootnote}{\arabic{footnote}}

\end{titlepage}

\newpage

\section{Introduction}

Recently, it was pointed out in \cite{ABJM} that the
three-dimensional ${\cal N}=6$ Chern-Simons theory
with  $U(N)_k \times U(N)_{-k}$ (ABJM theory)%
\footnote{Here $k$ and $-k$ denote the levels of Chern-Simons theory
with each $U(N)$.} has a holographic dual description in terms of
M-theory on $AdS_4 \times S^7/\mathbb{Z}_k$ or type IIA superstring
theory on $AdS_4 \times \mathbb{CP}^3$. After the discovery, there
has been remarkable progress in understanding AdS/CFT correspondence
\cite{Maldacena} in three dimensions, i.e., $AdS_4/CFT_3$
correspondence. Even in that situation, this duality is not yet
understood at the level of more familiar $AdS_5/CFT_4$
correspondence. One of the important aspects is how to add flavors
to the $AdS_4/CFT_3$ correspondence. In the ABJM theory we can add
flavor fields which belong to the fundamental representation of the
gauge group. Therefore the problem is how to reproduce the same
setup in the holographic dual theory.

This is important not only from the viewpoint of holographic duality
but also for the purpose of the application to some realistic models
in condensed matter physics. Indeed, the ABJM theory has been
employed to realize fractional quantum Hall effect (FQHE) recently
in \cite{FWRT} by considering edge states.\footnote{See also
\cite{Nakayama} for other possibilities of realization of FQHE.}
Notice also that in the standard Chern-Simons Ginzburg-Landau
description of FQHE, we need a charged scalar field. Such a field
cannot be found in the ABJM theory unless we introduce other fields
such as flavors.

The main purpose of this paper is to find a holographic description
of the ${\cal N}=3$ flavors in ABJM theory in terms of type IIA
superstring theory, motivated by a connection to FQHE. We will
concentrate on the case with a few number of flavors, therefore the
probe approximation is valid for large $N$ of gauge group $U(N)
\times U(N)$. This is also the case for describing flavors in ${\cal
N}=4$ super Yang-Mills gauge theory by adding probe D7-branes to
$AdS_5\times S^5$ \cite{KK}. We find that the flavor would be
introduced if we consider D6-branes wrapped over the Lens space
$S^3/{\mathbb Z}_2$ (or equivalently the real projective space
$\mathbb{RP}^3$) in $\mathbb{CP}^3$. This D6-brane has the degrees
of freedom of choosing a ${\mathbb Z}_2$ Wilson line and we will
identify this possibility with the choice of two gauge fields of
ABJM theory to which we add flavors. The probe D6-brane respects
desired R-symmetry and supersymmetry, and the same conclusion can be
derived from the argument based on the brane construction of ABJM
theory \cite{ABJM}. We will also calculate the fluctuations of the
D6-branes and observe that the conformal dimensions of dual
operators obtained from the analysis are consistent with what are
expected from the gauge theory side. In addition, we will give a
simple realization of FQHE in $AdS_4/CFT_3$ by adding the fractional
D2-branes to the ABJM theory. Adding flavor D6-branes to this setup,
we will give a realization of fractional quantum Hall plateau
transitions.

The organization of this paper is as follows; In section 2, we will
show how to introduce flavors to the ABJM theory and which D6-brane
embedding corresponds to the flavors. This is based on the analysis
of R-symmetry, supersymmetry and tyep IIB brane construction of ABJM
theory. In section 3, we perform the analysis of fluctuation
spectrum of the D6-branes. We will find the conformal dimensions of
the dual operators and confirm our identification of the flavor
D6-branes. In section 4, we will show how to model FQHE systems by
adding the fractional D2-branes to ABJM theory. Moreover, we will
realize fractional quantum Hall plateau transitions in the presence
of the flavor D6-brane. Section 5 reviews our conclusions and
suggests possibilities for future work.

\section{Flavor D6-branes in ABJM Theory}

The ${\cal N}=6$ supersymmetric Chern-Simons gauge theory in three
dimensions, often called ABJM theory, was shown to be dual to type
IIA superstring on $AdS_4\times \mathbb{CP}^3$ by taking the large
$N$ scaling limit with $\lambda=\f{N}{k}$ kept finite \cite{ABJM}.
This theory has the gauge group $U(N)\times U(N)$ with the level $k$
and $-k$, respectively. One of the most important deformations of
this theory should be adding flavors which belong to the fundamental
representation of the gauge groups. Since there are two gauge
groups, we expect two kinds of flavors. In the field theory side it
is straightforward to construct such a theory, and the purpose of
this section is to identify which configuration is the holographic
dual of the theory with flavors. We work within the probe brane
approximation, therefore the fundamental flavors are quenched. We
concentrate on the case preserving the maximal supersymmetry, that
is, the theory should possess ${\cal N}=3$ supersymmetry.

\subsection{ABJM Theory with Flavors}

First let us include flavors in the ABJM theory from the viewpoint
of gauge theory. The ABJM theory consists of two gauge multiplets
for the two copies of the gauge groups $U(N)\times U(N)$ and
bi-fundamental chiral fields $(A_1,A_2)$ in the $(N,\bar{N})$
representation and $(B_1,B_2)$ in the $(\bar{N},N)$ representation.
In order to add flavors to the ABJM theory, we would like to
introduce hypermultiplets while keeping ${\cal N}=3$ supersymmetry.
To achieve this, we add either of or both of the chiral multiplets
$(Q_1,\ti{Q}_1)$ and $(Q_2,\ti{Q}_2)$. Here
 chiral superfields $Q_1$ and $Q_2$
belong to $(N,{\bf 1})$ and $({\bf 1},N)$ representation,
respectively; and chiral superfields $\ti{Q_1}$ and $\ti{Q_2}$
belong to $(\bar{N},{\bf 1})$ and $({\bf 1},\bar{N})$, respectively.
For general discussions of the ${\cal N}=3$ Chern-Simons theory,
refer to e.g. \cite{GaXi,ABJM}.

In the ABJM theory, the interaction is essentially described by the superpotential
\be
W_{ABJM}=\mbox{Tr}[\vp_1^2-\vp_2^2]+\mbox{Tr}B_i\vp_{1}A_i+\mbox{Tr}A_i\vp_{2}B_i ~,
\ee
where $\vp_1$ and $\vp_2$ are the chiral superfields in the gauge multiplets.
We have added hypermultiplets to the ABJM theory, but the interaction
is determined by the requirement of ${\cal N}=3$ supersymmetry and it is given by
the following superpotential
\be
W_{flavor}=\mbox{Tr}\ti{Q}_1\vp_1 Q_1+\mbox{Tr}\ti{Q}_2\vp_2Q_2 ~.
\ee

Originally we have $SU(4)$ R-symmetry which rotates
$(A_1,A_2,\bar{B}_1,\bar{B}_2)$ in the ABJM theory. Even after the
flavors are added the theory still preserves ${\cal N}=3$
supersymmetry and the R-symmetry is now $SU(2)$, which acts on the
doublet $(A_i,\bar{B}_i)$ and $(Q_i,\ti{Q}_i)$. In addition, this
${\cal N}=3$ supersymmetric theory have an extra internal $SU(2)$
symmetry which acts on the doublets $(A_1,A_2)$ and $(B_1,B_2)$
simultaneously. Therefore the theory has the symmetry $SU(2)_R\times
SU(2)_I$.

In this way we have shown that one or two kinds of flavors
$(Q_1,\ti{Q}_1)$ and $(Q_2,\ti{Q}_2)$ can be introduced while
preserving ${\cal N}=3$ supersymmetry. Notice that mesonic operator
can be constructed as $\tilde Q_1 (AB)^l Q_1$ or $\tilde Q_1 (AB)^l
A Q_2$. If there is only one type of hypermultiplets, then we have
only the former type of mesonic operator. If there are both flavors,
we will have both operators. In the rest of this section we will see
how they are realized by adding flavor D6-branes in the holographic
dual geometry. In particular, the brane configuration realizing the
ABJM theory with flavors is constructed and the duality map confirms
the proposal.

\subsection{$AdS_4\times \mathbb{CP}^3$ Geometry}

It might be useful to start from the geometry dual to the ABJM
theory in order to identify the relevant D6-brane embedding. It was
argued in \cite{ABJM} that the dual theory is type IIA superstring
on $AdS_4\times \mathbb{CP}^3$,
whose metric is given by%
\footnote{In this paper we assume $\al=1$ and follow the notations
in \cite{Pol}.} \be ds^2=L^2(ds^2_{AdS_4}+4ds^2_{\mathbb{CP}^3}) ~,
\label{typeIIA} \ee where $L^2= R^3/(4k)$ with $R^6=2^5\pi^2 N k$.
The dilaton field is $e^{2\phi}= R^3 /k^3= 2^{\f{5}{2}}\pi \s{N
/k^5}$ and the background fluxes are \be F_2=\f{2k^2}{R^3}\omega
~,\ \ \   \ti{F}_4(\equiv F_4-C_1\we H_3)=-\f{3}{8}R^3 \ep_{AdS_4}
~,\ \ \ H_3=0 ~. \ee Here $\ep_{AdS_4}$ is the volume form of the
unit radius $AdS_4$ and $\omega$ is the K\"ahler form of
$\mathbb{CP}^3$. The metric of $\mathbb{CP}^3$ can be written down
explicitly as in \eqref{cp}, but it is instructive to construct the
metric for later purpose. This background preserves 24 out of total
32 supersymmetries of type IIA supergravity \cite{ABJM,BKKS}.

The metric of $\mathbb{CP}^3$ can be obtained by taking
large $k$ limit of the orbifold $S^7/\mathbb{Z}_k$.
Actually this fact was used to construct the dual
geometry \eqref{typeIIA} through the dimensional reduction of
the near horizon geometry of M2-branes at the orbifold
${\mathbb C}^4/{\mathbb Z}_k$, which is given by
$AdS_4\times S^7/{\mathbb Z}_k$.
We can express $S^7$  by the complex
coordinates $X_1,X_2,X_3$ and $X_4$ with the constraint
$|X_1|^2+|X_2|^2+|X_3|^2+|X_4|^2=1$.
It is convenient to parameterize $S^7$ as%
\footnote{We follow the notation in \cite{NTP}.}
\begin{align}
X_1&=\cos\xi \, \cos\f{\theta_1}{2} \, e^{i\f{\chi_1+\vp_1}{2}} ~, \qquad
X_2=\cos\xi \,  \sin\f{\theta_1}{2} \, e^{i\f{\chi_1-\vp_1}{2}} ~, \nonumber \\
X_3&=\sin\xi \,  \cos\f{\theta_2}{2} \, e^{i\f{\chi_2+\vp_2}{2}} ~, \qquad
X_4=\sin\xi \, \sin\f{\theta_2}{2} \, e^{i\f{\chi_2-\vp_2}{2}} ~,
\label{angles}
\end{align}
where the ranges of the angular variables are $0\leq \xi
<\f{\pi}{2}$, $0\leq \chi_i <4\pi$, $0\leq \vp_i < 2\pi$ and $0\leq
\theta_i<\pi$. The $\mathbb{Z}_k$ orbifold action is taken along the
$y$-direction as $y\sim y+\f{2\pi}{k}$, where the new coordinate $y$
is defined by
\begin{align}
\chi_1=2y+\psi ~, \qquad \chi_2=2y-\psi ~.
\end{align}
In the new coordinate system, the $S^7$ can be
rewritten as
\be
ds_{S^7}^2=ds^2_{\mathbb{CP}^3}+(dy+A)^2,
\ee where
\begin{align}
A=\f{1}{2}(\cos^2\xi-\sin^2\xi)d\psi +\f{1}{2}\cos^2\xi
\cos\theta_1 d\vp_1+ \f{1}{2}\sin^2\xi \cos\theta_2 d\vp_2 ~.
\end{align}
In this way we find the metric of $\mathbb{CP}^3$ as
\begin{align} \label{cp}
ds^2_{\mathbb{CP}^3}&=d\xi^2+\cos\xi^2\sin^2\xi\left(d\psi+\f{\cos\theta_1}{2}d\vp_1-
\f{\cos\theta_2}{2}d\vp_2\right)^2 \\ & \qquad
+\f{1}{4}\cos^2\xi\left(d\theta_1^2+\sin^2\theta_1
d\vp_1^2\right)+\f{1}{4}\sin^2\xi(d\theta_2^2+\sin^2\theta_2
d\vp_2^2) ~. \nonumber
\end{align}
The ranges of the angular valuables are given by \be 0\leq \xi
<\f{\pi}{2} ~,\ \ \ 0\leq \psi < 2\pi ~, \ \ \ 0\leq \theta_i<\pi
~,\ \ \ \ 0\leq \vp_i \leq 2\pi ~. \label{range} \ee In this
coordinate system, the RR 2-form $F_2=dC_1$ in the type IIA string
is explicitly given by \ba F_2&=& k\Bigl(-\cos\xi\sin\xi d\xi \we
(2d\psi+\cos\theta_1d\vp_1-\cos\theta_2 d\vp_2)\no &&
-\f{1}{2}\cos^2\xi\sin\theta_1 d\theta_1\we d\vp_1
-\f{1}{2}\sin^2\xi\sin\theta_2 d\theta_2 \we d\vp_2\Bigr) \equiv
-\f{2k^2}{R^3}\omega ~. \ea The explicit expression of the K\"ahler
form $\omega$ can also be read off from this equation.

\subsection{Flavor D6-branes}
\label{dsix}

Utilizing the explicit metric of $\mathbb{CP}^3$, we would like to
discuss the D-brane configuration in $AdS_4\times \mathbb{CP}^3$,
which is dual to adding flavors to ABJM theory. The corresponding
D6-brane should be wrapped over
$AdS_4$ times a topologically trivial%
\footnote{If a brane is wrapped over a topologically trivial cycle
in $\mathbb{CP}^3$, then the brane over $AdS_4$ does not carry any
charge. Otherwise, it is not possible to wrap the whole $AdS_4$
space. The topologically trivial cycle tends to shrink due to the
brane tension, but this brane configuration can be actually
stabilized due to the curvature of AdS space. See, e.g., \cite{KK}
for more detail.} 3-cycle in $\mathbb{CP}^3$. As discussed above,
the original ABJM theory has $SU(4)$ R-symmetry, which corresponds
to the $SU(4)$ symmetry of $\mathbb{CP}^3$. Adding the flavors
reduces the R-symmetry into $SO(4) = SU(2)_R \times SU(2)_I$,
therefore we should find a cycle with the $SO(4)$ symmetry.

We assume that the coordinates $(X_1,X_2,X_3,X_4)$ correspond to $(A_1,\bar{B}_1,\bar{B}_2,A_2)$, and in this case
the $SU(2)$ symmetry rotates $(X_1,X_2)$ and $(X_3,X_4)$ at the same time.
We want to have a 3-cycle invariant under this rotation,
and a natural one is given by
\begin{align}
\theta_1=\theta_2(=\theta) ~,\ \ \vp_1=-\vp_2(=\vp),\ \ \ \xi=\f{\pi}{4} ~.
\label{6cond}
\end{align}
The induced metric becomes \be
ds^2=4L^2\left(\f{1}{4}(d\psi+\cos\theta
d\vp)^2+\f{1}{4}(d\theta^2+\sin^2\theta d\vp^2)\right)
~,\label{thsp} \ee where $0\leq \psi<2\pi$, $0\leq \theta<\pi$ and
$0\leq \vp<2\pi$. This metric looks like the metric of
a regular $S^3$ with the unit radius,%
\footnote{ Notice that  we can rewrite (\ref{thsp}) as $ds^2=
d\xi^2+\cos^2\xi d\phi_1^2+\sin^2\xi d\phi_2^2$ by setting $\xi =
\theta/2$, $\psi=\phi_1+\phi_2$ and $\vp=\phi_1-\phi_2$. } though in
that case the periodicity should be $0\leq \psi<4\pi$ instead of
$0\leq \psi<2\pi$. Therefore we conclude that the 3-cycle we found
is actually the Lens space $S^3/{\mathbb Z}_2$. If we describe the
$S^3$ by $w_1^2+w_2^2+w_3^2+w_4^2=1$, then the ${\mathbb Z}_2$
orbifold action is given by \be (w_1,w_2,w_3,w_4)\to
(-w_1,-w_2,-w_3,-w_4) ~. \ee An important property is that this
${\mathbb Z}_2$ action is the center of $SO(4)$, and hence a
D6-brane wrapped over this $S^3/{\mathbb Z}_2$ preserves the $SO(4)$
symmetry as expected.
 The volume of $S^3/{\mathbb Z}_2$ can be computed as
 Vol$(S^3/{\mathbb Z}_2)=8\pi^2L^3$. As we will see later, this $S^3/{\mathbb Z}_2$ is the same as the
$\mathbb{RP}^3$ which is embedded into $\mathbb{CP}^3$ in a rather trivial way.

It is important to notice that there is a non-trivial torsion
cohomology as \be H^2(S^3/{\mathbb Z}_2,\mathbb{Z})={\mathbb Z}_2 ~,
\ee and a gauge theory on this manifold has two vacua due to the
$\mathbb{Z}_2$ torsion. Let us define $[\ap]$ as the torsion 1-cycle
in $S^3/{\mathbb Z}_2$ generated by $0\leq \psi<2\pi$, then the
${\mathbb Z}_2$ charge is interpreted as the ${\mathbb Z}_2$ Wilson
loop \be e^{i\int_{[\ap]} A}=\pm 1 ~. \ee In other words, we can
construct two types of D6-branes depending on the Wilson loop. In
the following, we will show that one of them provides a flavor for
one of the two $U(N)$ gauge groups and the other does the other one.
This is motivated by the Douglas-Moore prescription of D-branes at
${\mathbb Z}_2$ orbifold \cite{DoMo}, although our setup is a T-dual
of them.

\subsection{Supersymmetry of D6-brane}

In the previous subsection we have found a candidate 3-cycle on
which the flavor D6-brane should be wrapped based on the R-symmetry
argument. Since a three-dimensional ${\cal N}=3$ superconformal
field theory possesses 12 supersymmetries, our flavor brane
configuration should preserve half of the 24 supersymmetries in the
bulk. In order to count the number of supersymmetries of the
D6-brane configuration in $AdS_4 \times \mathbb{CP}^3$, it is useful
to uplift to M-theory and to utilize the Killing spinor in
$AdS_4\times S^7/\mathbb{Z}_{k}$. Adding the eleventh dimension $y$,
the two ten-dimensional 16-components (Weyl) Killing spinors $\{
\epsilon_{\pm}\}$ is combined into a 11D 32-component Killing
spinor. Following \cite{NiTa} we define the angular coordinates
$X_i=\mu_i e^{i\zeta_i}$ instead of (\ref{angles}) with
$\{\mu_1,\mu_2,\mu_3,\mu_4\}=\{\sin{\alpha},\cos{\alpha}\sin{\beta},
\cos{\alpha}\cos{\beta}\sin{\gamma},\cos{\alpha}\cos{\beta}\cos{\gamma}\}$.
The Killing spinor is now given by
\begin{equation}
\epsilon=e^{\frac{\alpha}{2}\hat{\gamma}\gamma_4}
e^{\frac{\beta}{2}\hat{\gamma}\gamma_5}
e^{\frac{\gamma}{2}\hat{\gamma}\gamma_6}
e^{\frac{\xi_1}{2}\gamma_{47}} e^{\frac{\xi_2}{2}\gamma_{58}}
e^{\frac{\xi_3}{2}\gamma_{69}}
e^{\frac{\xi_4}{2}\hat{\gamma}\gamma_{10}}
e^{\frac{\rho}{2}\hat{\gamma}\gamma_{1}}
e^{\frac{t}{2}\hat{\gamma}\gamma_{0}}
e^{\frac{\theta}{2}\gamma_{12}} e^{\frac{\phi}{2}\gamma_{23}}
\epsilon_0 ~ ,
\end{equation}
where
$(x^0,x^1,\ddd,x^{10})=(t,r,\theta,\phi,\alpha,\beta,\gamma,\xi_1,\xi_2,\xi_3,\xi_4)$
and $\epsilon_0$ is a constant 32-component Majorana spinor in 11D.
The eleventh dimension $y$ is a linear combination of the four
phases $\{\zeta_i\}$.

In $AdS_4\times \mathbb{CP}^3$, consider a D6-brane extending along
the entire $AdS_4$ and the $\{\alpha, \beta, \gamma\}$-directions
while sitting at constant phase directions. When lifted to M-theory,
it corresponds to a Taub-NUT spacetime along the
$016789$-directions. Then the supersymmetries preserved are given by
the constraint \ba \Gamma_{6}\ep=\ep \qquad \textrm{where} \qquad
\Gamma_{6}=\gamma_{0123456} ~. \ea Therefore it projects out half of
the supersymmetries by \be \gamma_{0123456}\epsilon_0=\epsilon_0
~.\ee Then the orbifolding
 action $z_i\rightarrow z_i e^{i2\pi/k}$ further projects out $4$
supersymmetries. In total, this 11D system with the Taub-NUT
spacetime has $12$ supersymmetries. Performing the dimensional
reduction on the $y$-direction, we return to the D6-brane extending
along $AdS_4$ and the $\{\alpha,\beta,\gamma\}$-directions inside
$\mathbb{CP}^3$ and it preserves 12 supersymmetries. Since
$\{\alpha,\beta,\gamma\}$ are the three real directions in
$\mathbb{CP}^3$, the 3-cycle wrapped by the D6-brane is
$\mathbb{RP}^3$.

Utilizing the $SU(4)$ symmetry of  $\mathbb{CP}^3$, we can show that
the above D6-brane configuration is indeed the one obtained before.
We perform the following $SU(4)$ symmetry transformation of
$\mathbb{CP}^3$
\begin{equation}
\f{1}{\s{2}}(X_1+X_3) \to X_1 ~, \ \ \ \f{-i}{\s{2}}(X_1-X_3) \to
X_2 ~, \ \ \ \f{1}{\s{2}}(X_2+X_4)  \to X_3~,\ \ \
\f{i}{\s{2}}(X_2-X_4) \to X_4~. \end{equation} Then, the cycle
defined by the condition \eqref{6cond} is mapped to the one with the
induced metric
\begin{equation} ds^2=d\xi^2+\f{1}{4}\cos^2\xi
d\theta_1^2+\f{1}{4}\sin^2\xi d\theta_2^2 ~, \label{thspp}
\end{equation}
which may be given by the replacement
\begin{equation} \psi+\varphi
\to \theta_1 ~,\ \ \ -\psi+\varphi \to \theta_2~,\ \  \
\f{\theta}{2} \to \xi ~.
\end{equation}
This cycle can also be obtained by setting $\vp_i=0$ and $\chi_i=0$
in the new coordinates (\ref{angles}). If we take into account the
presence of $\mathbb{Z}_k$ orbifold action carefully, we can find
that the ranges of coordinates are $0\leq \xi <\pi/2$ and $0\leq
\theta_i< 2\pi$ with the ${\mathbb Z}_2$ identification $\theta_1\to
\theta_1+\pi$ and $\theta_2\to \theta_2+\pi$. Thus we again obtain
$S^3/{\mathbb Z}_2$, which can be mapped to the previous one
(\ref{thsp}) by the $SU(4)$ symmetry of the background. In this way
we have proved that the D6-brane over $S^3/{\mathbb Z}_2$ discussed
in the previous subsection preserves 12 supersymmetries as we wanted
to show. Notice also that from this construction we can clearly
understand the 3-cycle $S^3/{\mathbb Z}_2$ as the  $\mathbb{RP}^3$
inside $\mathbb{CP}^3$.

\subsection{Relation to Brane Configuration}
\label{Brane}

One of the confirmation of the duality between the ABJM theory and
type IIA superstring on $AdS_4\times \mathbb{CP}^3$ is made through
the realization of ABJM theory with the type IIB brane configuration
\cite{ABJM} (see also \cite{Ohta,Ber}). Therefore, the construction
of brane configuration corresponding to the ABJM theory with flavors
would give a strong support of our identification of flavor D-brane.

Let us begin with the ABJM theory without flavor. We introduce a
standard cartesian coordinate $x^0,x^1,\ddd,x^9$ with $x^6$
compactified on a small circle. Then the type IIB brane
configuration of the ABJM theory is given by $N$ D3-branes which
extend in the $0126$-directions, a NS5-brane in $012345$ and a
$(1,k)$5-brane in $012[3,7]_\theta [4,8]_\theta [5,9]_{\theta}$.
Here $[i,j]_\theta$ means that it extends in the particular
direction between $\de_i$ and $\de_j$ so that it preserves ${\cal
N}=3$ supersymmetry \cite{Ohta,Ber}. Since the NS5-brane and
$(1,k)$5-brane divide the circular D3-branes into two segments, the
gauge group becomes $U(N)\times U(N)$. The chiral matter multiplets
$A_i$ and $B_i$ come from open strings between these two parts of
the D3-branes.

In order to introduce flavors to ABJM theory, we need to insert
D5-branes to this setup. Here we introduce D5-branes in the
$012789$-directions such that the brane configuration preserves
${\cal N}=3$ supersymmetry as confirmed in appendix A.\footnote{ The
same D5-brane is also discussed in the IIB brane configuration in
the independent work \cite{Niarchos}, quite recently from a
different motivation.} Since there are two segments of D3-branes, we
can insert D5-branes in either or both of these two segments. If we
insert a D5-brane in a segment, then we have one type of
hypermultiplets, say, $(Q_1, \tilde Q_1)$. If we insert another
D5-brane in the other segment, then we have one more type of
hypermultiplets $(Q_2, \tilde Q_2)$. When $N^f_1$ and $N^f_2$
D5-branes are inserted in each segments of D3-branes, the number of
flavors for $Q_1$ and $Q_2$ are increased by $N^f_1$ and $N^f_2$,
respectively. In this paper we will set $N^f_1$ and $N^f_2$ to be
zero or one. Mesonic operators may be interpreted as the strings
stretching between the D5-branes, and strings between the same brane
correspond to the type of $\tilde Q_1 (AB)^l Q_1$ and strings
between the different branes correspond to the type of
 $\tilde Q_1 (AB)^l A Q_2$. Notice that this D5-brane is a standard
flavor D-brane in the brane configurations of three-dimensional
${\cal N}=4$ supersymmetric Yang-Mills gauge theory \cite{HaWi}.

In the following we will show that when mapped to type IIA theory
the above D5-brane actually corresponds to the D6-brane wrapped over
$AdS_4$ times the cycle $S^3/{\mathbb Z}_2$ (\ref{thsp}) obtained
above. We start with the case without flavor again. Via the standard
duality map, the type IIB brane configuration can be lifted to
M-theory with M2-branes at the intersection of two KK monopoles.
Before adding the M2-branes, this geometry takes the form of
$R^{1,2}\times X_8$ and the explicit metric of $X_8$ can be found in
\cite{ABJM} (see also \cite{GGPT}). There the coordinates of
eight-dimensional manifold $X_8$ were expressed by
$(\vp_1,\vec{x}^1)$, $(\vp_2,\vec{x}^2)$, which is essentially two
copies of Taub-NUT spacetimes warped with each other.  The relation
between this coordinate of $X_8$ and the brane configuration is
given by $\vp_1=x^6$, $\vp_2=x^{10}$, $\vec{x}^1=(x^7,x^8,x^9)$ and
$\vec{x}^2=(x^3,x^4,x^5)$. Comparing with the coordinates
$(X^1,X^2,X^3,X^4)$ of $\mathbb{C}^4/\mathbb{Z}_k$ in
(\ref{angles}), we have
\begin{align}
& \chi_1=-2\vp'_1\equiv-2(\vp_1-\f{\vp_2}{k}) ~,\ \ \ \chi_2=\vp'_2\equiv\f{\vp_2}{k} ~, \label{changea}\\
& \vec{x}'_1=\vec{x}_1=r^2\cos^2\xi(\cos\theta_1,\sin\theta_1\cos\vp_1,\sin\theta_1\sin\vp_1) ~,\ \ \no
& \vec{x}'_2=\vec{x}_1+k\vec{x}_2=r^2\sin^2\xi(\cos\theta_2,\sin\theta_2\cos\vp_2,\sin\theta_2\sin\vp_2) ~,
\nonumber
\end{align}
where $r$ is defined by $\sum_{i=1}^4|X_i|^2=r^2$.
This leads to $x^6=\psi$ and $x^{10}=ky-\f{k}{2}\psi$.

We would like to introduce a 6-brane in this setup. Our D6-brane
wrapped over $S^3/{\mathbb Z}_2$ (\ref{thsp}) corresponds to a
KK-monopole in M-theory. As is clear from the description in the
coordinate (\ref{angles}), it is simply expressed as the
codimension-three surface of $\vec{x}'_1=\vec{x}'_2$, which leads to
the constraint $\vec{x}_2=0$. Taking the T-duality in the
$6$-direction (notice that the D6-brane extends in the
$6$-direction), it becomes a D5-brane in the IIB string which
extends in the $012789$-directions. This argument almost confirmed
that our D6-brane over $S^3/{\mathbb Z}_2$ corresponds to the
D5-brane introduced in the type IIB brane configuration with one
subtlety. Namely, we only have to explain the fact that a D5-brane
can be inserted in either of the two segments. Actually this fact is
consistent with our D6-brane setup since we have the choice of
${\mathbb Z}_2$ Wilson loop in the $\psi$-direction. Performing the
T-duality in the $x^6=\psi$ direction, these two possibilities
correspond to the two segments of the D3-branes on which we can
place the D5-brane.

\section{Meson Spectrum from Flavor Brane}

One of the most important checks of AdS/CFT correspondence is the
comparison of spectrum. In this section we would like to investigate
the fluctuation of D6-brane wrapping over $AdS_4 \times S^3/{\mathbb
Z}_2$ inside $AdS_4 \times \mathbb{CP}^3$. The spectrum of the
fluctuation should be reproduced by the conformal dimensions of dual
operators. We will study the fluctuation of a scalar mode transverse
to the $S^3/{\mathbb Z}_2$ in $\mathbb{CP}^3$ and the gauge field on
the worldvolume. We start from the D6-brane action
\begin{align}\label{D6action}
 S_{D6} &= - \frac{1}{(2 \pi)^6}
  \int d^{1 + 6} x e^{- \phi} \sqrt{ - \det ( g_{ab} + 2 \pi F_{ab}) }
  \\ &\qquad\qquad\qquad\qquad
          + \frac{(2 \pi)^2}{2 (2 \pi)^6}  \int C_{3} \wedge F \wedge F
          + \frac{1}{ (2 \pi)^6}  \int C_{7}~.
           \nonumber
\end{align}
Here $g_{ab}$ is the induced metric of the D6-brane, $F_{ab}$ is the
field strength on the worldvolume, and $C_3$, $C_7$ are the induced
3-form and 7-form potentials. There are other types of Chern-Simons
term, but we included only those relevant for our purpose. We adopt
the static gauge and the measure is $d^{1+6}x = d t d x d y d r d
\theta d \psi d \varphi$. In the following we use $\mu$ for $t,x,y$
and $i,j$ for $S^3$ coordinates. In the $i,j$ label directions are
given by
\begin{align}
 d \sigma^1 = \frac{1}{2} d \theta ~, \qquad
 d \sigma^2 = \frac{1}{2} d \psi ~, \qquad
 d \sigma^3 = \frac{1}{2}  \sin \theta d \varphi ~. \qquad
\end{align}
In the case of D7-brane in $AdS_5 \times S^5$ similar analyses have
been done in \cite{KMMW} (see also \cite{KK}).

\subsection{Scalar perturbation}

First we study the fluctuations of a scalar mode, which correspond
to the scalar perturbation of D6-brane orthogonal to the worldvolume
directions. Here we consider only the fluctuation of $\xi = \pi/4 +
\eta$ with small $\eta$ and the fluctuations along the other two
directions will be obtained by the symmetry argument. For this
purpose the Chern-Simons term with 7-form potential is important.
Using the fact that $F_2=*F_8=-\frac{2k^2}{R^3}\omega$, the 7-form
potential $C_7$ can be written as
\begin{align}
C_7=-\frac{k^2L^4}{R^3}\sigma\wedge \omega\wedge r^2 dt\wedge dx\wedge dy\wedge dr ~,
\end{align}
where $\sigma$ is defined by $d\sigma=\omega$.
Under the condition of $\theta_1=\theta_2$
and $\varphi_1+\varphi_2=0$, we find
\begin{align}
& \sigma=-L^2(\cos^2\xi-\sin^2\xi)(d\psi+\cos\theta d\varphi) ~, \\ \nonumber
& \omega=L^2\left(4\cos\xi\sin\xi d\xi\wedge d\psi+(\cos^2\xi-\sin^2\xi)\sin\theta d\theta\wedge d\varphi
+4\cos\xi\sin\xi\cos\theta d\xi\wedge d\varphi\right) ~, \nonumber
 \end{align}
therefore we have
\begin{align}
C_7=\frac{k^2L^8}{R^3}(\cos^2\xi-\sin^2\xi)^2 r^2\sin\theta dt\wedge dx\wedge dy\wedge dr\wedge d\psi\wedge d\theta\wedge d\varphi ~.
\end{align}

Expanding the D6-brane action \eqref{D6action},
the quadratic term of $\eta$ is given by
\begin{align}
 \delta S = \frac{k }{2(2 \pi)^6 L} \int d^{1 + 6} x \sqrt{- \det g}
( 2 g^{ab} \partial_a \eta \partial_b \eta
- 4 \eta^2 ) ~.
\end{align}
In our notation $\sqrt{- \det g} = L^7 r^2 \sin \theta$.
The equation of motion for $\eta$ leads to
\begin{align}
 \frac{1}{\sqrt{- \det g}} \partial_a ( \sqrt{-\det g} g^{ab} \partial_b ) \eta
  + 2 \eta = 0 ~,
\end{align}
therefore we have
\begin{align}
  \quad \frac{1}{r^{2}}
 \partial_\mu \partial^\mu \eta + \frac{1}{r^{2}} \partial_r (r^4 \partial_r )\eta
 + \tfrac14 D_i D^i \eta + 2 \eta = 0 ~.
\end{align}
Here $D_i$ represents covariant derivatives on $S^3$.
Using the separation of variables we can write as
\begin{align}
 \eta = \rho (r) e^{i k \cdot x} Y^{l} (S^3)
\end{align}
with the spherical harmonics
\begin{align}
 D_i D^i  Y^{l}(S^3) = - l (l+2)  Y^{l}(S^3) ~ .
\end{align}
As discussed in the apppendix \ref{orbifold}, the ${\mathbb Z}_2$
orbifold action restricts $l \in 2 {\mathbb Z}$. If the scalar field
feels the ${\mathbb Z}_2$ holonomy along the $\psi$-direction, then
the restriction is $l \in 2 {\mathbb Z} + 1$. With the help of
separation of variables, the equation of motion reduces to
\begin{align}
 \left [ \partial_r^2 + \frac{4}{r} \partial_r
  + \frac{8 - l(l+2)}{4 r^2}  - \frac{k^2}{r^4} \right] \rho (r) = 0 ~.
\end{align}
We assume the regularity at the horizon of $AdS_4$, i.e. at $r = 0$.
Then the above equation can be solved by
the modified Bessel function as
\begin{align}
 \rho (r) = r^{-\frac{3}{2}} K_{\frac{l+1}{2}} (\tfrac{k}{r}) ~.
\end{align}
As usual the conformal dimension of dual operator $\Delta$ can be
read off from the boundary behavior at $r \to \infty$ as $r^{-
\Delta}$ or $r^{3 - \Delta}$. Expanding the solution around $r \to
\infty$, we find
\begin{align}
 \rho (r) \sim c_1 r^{- \frac{3}{2} + \frac{l+1}{2}} + c_2
  r^{- \frac{3}{2} - \frac{l+1}{2}} ~,
\end{align}
with some coefficients $c_1,c_2$. Therefore the conformal
dimension of dual operator is
\begin{align}
  \Delta = \frac{l}{2} + 2 ~.
\end{align}
Without any holonomy the conformal dimension is
$\Delta = 2 + n$ with $n = 0,1,2,\cdots$ and the dual operator
is of the form $\tilde \psi_1 (AB)^{n} \psi_1$. The lowest one $n=0$ is interpreted
as the (supersymmetric) mass deformation of the flavor.
In the case with ${\mathbb Z}_2$ holonomy,
the conformal dimension is
$\Delta = 2 + n + 1/2$ with $n = 0,1,2,\cdots$ and the dual operator
is of the form $\tilde \psi_1 (AB)^{n} A \psi_2$.%
\footnote{Since the other two scalar modes can be obtained by the
symmetry transformation of $\mathbb{CP}^3$ while fixing the
$S^3/\mathbb{Z}_2$, it is natural to guess that the conformal
dimensions of their dual operators are also the same as in this
case.}

\subsection{Vector perturbation}

On a D6-brane, there is a $U(1)$ gauge field and we can study the
spectrum due to a small shift of the gauge field. The equations of
motion are given by
\begin{align}
 \frac{1}{ \sqrt{- \det g}}
 \partial_a ( \sqrt{ - \det g} F^{ab}) - \frac{3}{8}
\epsilon^{bij} \partial_i A_j = 0~.
\label{vectoreom}
\end{align}
Here we have used the fact that the induced 3-form potential can be
written as
\begin{align}
 C_3 = - \frac{k L^2}{2} r^3 dt \wedge dx \wedge dy ~.
\end{align}
Following \cite{KMMW} it can be shown that it is enough to consider
the following three types of gauge field configuration. They are
given by
\begin{align}
 &\text{Type I:} \qquad
  A_\mu = 0 ~, \qquad A_r = 0 ~, \qquad
  A_i = \rho^{\pm}_I (r) e^{ik \cdot x} Y^{l,\pm}_i (S^3) ~, \\
 &\text{Type II:} \quad
  A_\mu = \xi_\mu \rho_{II} (r) e^{ik \cdot x} Y^{l} (S^3)~, \quad
   \xi \cdot k = 0 ~, \quad
  A_r = 0 ~, \quad
  A_i = 0 ~, \\
 & \text{Type III:} \quad
  A_\mu = 0 ~,
  \quad A_r = \rho_{III} (r) e^{ik \cdot x} Y^{l} (S^3) ~, \quad
  A_i = \tilde \rho_{III} (r) e^{ik \cdot x} D_i Y^{l} (S^3) ~.
\end{align}
The vector components along $S^3$ directions can be expanded by
the vector spherical harmonics, which satisfy
\begin{align}
 &D_i D^i Y^{l,\pm}_j -R^k_j Y^{l,\pm}_k
   = - (l+1)^2 Y^{l,\pm}_j ~,  \\
   &\epsilon_{ijk} D_j Y^{l,\pm}_k
    = \pm ( l+1) Y^{l,\pm}_k , \quad
   D^i Y^{l,\pm}_i = 0 ~, \nonumber
\end{align}
where $R^k_j = 2 \delta^k_j$ is the Ricci tensor for $S^3$
with the unit radius.
They belong to $(\frac{l \mp 1}{2}, \frac{l \pm 1}{2})$
representation with respect to $SU(2)_R \times SU(2)_L$.

Let us start with type I case. The equations of motion
\eqref{vectoreom} lead to
\begin{align}
 \frac{1}{r^2} \partial_\mu \partial^{\mu} A^j +
 \frac{1}{r^2} \partial_r ( r^4 \partial_r A^j)
  + \frac{1}{4} (D_i D^i A^j - 2 \delta^j_i A^i)
  - \frac{3}{2} \epsilon_{jkl} \partial_k A^l = 0 ~.
\end{align}
Here we have used
\begin{align}
 \frac{1}{\sqrt{g}} \partial_i \sqrt{g}
 (\partial^i A^j - \partial^j A^i)
  = D_i D^i A^j + D^j D_i A^i - [D_i , D^j] A^i
\end{align}
and $D_i A^i = 0$, $[D_i , D^j] A^i = R_{i}^{~j} A^i$.
Then we find
\begin{align}
  \partial_r^2 \rho_I^{\pm} + 4 \partial_r \rho_I^{\pm}
 - \frac{k^2}{r^4} \rho^{\pm}_I
 - \frac{(l+1)^2}{4 r^2} \rho^{\pm}_I
 \mp \frac{6}{4} (l+1) \rho^{\pm}_I = 0 ~.
\end{align}
The solution for $\rho^+_I (r)$ regular at $r = 0$ is
\begin{align}
 \rho^+_I (r) = r^{- \frac{3}{2}} K_{\frac{l+4}{2}} (\tfrac{k}{r})
   \sim c_1 r^{- \frac{1}{2} (l+7)} + c_2 r^{ \frac{1}{2} (l+1)} ~,
\end{align}
thus the conformal weight of the dual operator is
 $\Delta_+ = \frac{l}{2} + \frac{7}{2}$, where
$l \in 2 {\mathbb Z} + 1$ without holonomy and
$l \in 2 {\mathbb Z} $ with ${\mathbb Z}_2$ holonomy.
The solution for $\rho^-_I (r)$ regular at $r = 0$  is
\begin{align}
 \rho^-_I (r) = r^{- \frac{3}{2}} K_{\frac{l-2}{2}} (\tfrac{k}{r})
   \sim c_1 r^{- \frac{1}{2} (l+1)} + c_2 r^{ \frac{1}{2} (l - 5)} ~,
\end{align}
thus the conformal weight of the dual operator is
 $\Delta_- = \frac{l}{2} + \frac{1}{2}$
with the same condition for $l$ as in $\rho^+_{I}$ case. The lowest
one is given by $l = 1$ case, which is in the $(1,0)$ representation
and transforms as the triplet of $SU(2)_R$. The dual operator can be
identified with the triplet ${\cal O}^1
=\{\bar{Q}Q-\bar{\ti{Q}}\ti{Q},\ti{Q}Q,\bar{\ti{Q}}\bar{Q}\}$ of the
scalar field in the hypermultiplet. This identification is quite
important since the other cases follow only with the (super)symmetry
arguments.

The type II case can be analyzed in the same way. The equations of
motion lead to
\begin{align}
 \frac{1}{r^4} \partial_\nu \partial^\nu A_\mu
  +  \partial_r \left(r^2 \partial_r \left(\frac{1}{r^2} A_\mu \right) \right)
  + \frac{1}{4 r^2} D_i D^i A_\mu = 0 ~,
\end{align}
thus we obtain
\begin{align}
 \left( - \frac{k^2}{r^4} + \partial_r^2 + \frac{2}{r} \partial_r
  + \frac{8 - l (l+2)}{4 r^2}  \right) \rho_{II} (r) = 0 ~.
\end{align}
The solution to this equation regular at $r=0$ is given by
\begin{align}
 \rho_{II} (r) = r^{- \frac32} K_{\frac{l+1}{2}} (\tfrac{k}{r})
  \sim c_1 r^{\frac{l}{2} - 1} + c_2 r^{ - \frac{l}{2} - 2} ~,
\end{align}
and the conformal dimension of the dual operator is $\Delta =
\frac{l}{2} + 2$. The restriction to $l$ is the same as the scalar
case and $l \in 2 {\mathbb Z}$ without holonomy and $l \in 2
{\mathbb Z} + 1$ with ${\mathbb Z}_2$ holonomy.

For type III case we first set $b = \mu$. Then we obtain the
relation
\begin{align}
 \partial_r (r^2 \rho_{III} (r) )
  - \tfrac{1}{4} l(l+2) \tilde \rho_{III} (r) = 0 ~.
\end{align}
For $l = 0$, the solution behaves as $\rho_{III} \sim 1/r$. Since it
is singular at $r =0$ we remove $ l=0$ mode. Then the equations of
motion for $b = r$ or $b = j$ read
\begin{align}
 r^2 \partial_r^2 (r^2 \rho_{III} (r)) - k^2 \rho_{III} (r)
 - \tfrac{1}{4} r^2 l ( l+2 ) \rho_{III} (r) = 0 ~,
\end{align}
and the solution regular at $r=0$ is
\begin{align}
 \rho_{III} (x) = r^{- \frac{3}{2}} K _{\frac{l+1}{2}} (\tfrac{k}{r})
 \sim c_1 r^{-\frac{l}{2} - 2 } + c_2 r^{ \frac{l}{2} - 1 } ~,
\end{align}
thus $\Delta = \frac{l}{2} + 2 $, where $ l = 2,4,6, \cdots $
without holonomy and $ l = 1,3,5,\cdots$ with ${\mathbb Z}_2$
holonomy.

\subsection{Spectrum and the ${\mathbb Z}_2$ Wilson Loop}

Let us summarize the results obtained in this section. Due to the
choice of the ${\mathbb Z}_2$ Wilson loop, we have two types of
D6-branes. Irrespective of the choice of Wilson loop, open strings
on the same brane do not feel the effects of Wilson loop. Therefore,
the scalar fields and the gauge field from the open string do not
receive the $\mathbb{Z}_2$ holonomy. The conformal dimensions of the
dual operators are in this case $\Delta = n + 2$ with $n =
0,1,2,\cdots$ for scalar fields and gauge field in the $(n,n)$
representation. For type III case $n=0$ is removed. For type I, it
is given by $\Delta = n + 3$ in the $(n-1,n)$ representation and
$\Delta = n + 1$ in the $(n + 1,n)$ representation. Notice that the
conformal dimensions are always integers. This is consistent with
our identification of a D6-brane with a flavor for either of the two
$U(N)$ gauge groups, where excitations of bi-fundamental scalars
$A_i$ and $B_i$ should always include even number of these scalar
fields with $\Delta = 1/2$ as explained before.

On the other hand, if we consider an open string between
two different branes, then the fields coming from the
open string receives the $\mathbb{Z}_2$ holonomy along the
non-trivial cycle. In this case
the conformal dimensions of dual operator are
$\Delta = n + 5/2$ with $n = 0,1,2,\cdots$ for scalar fields
and gauge field in the $(n+1/2,n+1/2)$ representation.
For type I, it is given by $\Delta = n + 7/2$ in the
$(n-1/2,n+1/2)$ representation and  $\Delta = n + 3/2$ in the
$(n + 3/2,n+1/2)$ representation. The conformal dimensions
take always half integer numbers in this case.
Again these facts can be explained if we assume that two D6-branes with
different ${\mathbb Z}_2$ Wilson lines correspond to flavor for two different gauge groups. In this way,
our results of the spectrum support our identification of
flavor D6-branes described in section \ref{dsix}.

\section{Fractional Quantum Hall Effect and ABJM Theory}

One interesting application of ABJM theory to condensed matter
physics is to use it to model fractional quantum Hall effects
holographically \cite{FWRT}. This
stems from the fact that the low energy
effective description of FQHE with filling faction $\nu=\f{1}{k}$
(where $k\in \mathbb{Z}$) can be captured by a $U(1)_k$ Chern-Simons
theory (see e.g. the text book \cite{Wen}). The action is
\begin{equation}
S_{FQHE}=\f{k}{4\pi}\int A\we
dA+\f{1}{2\pi}\int A\we F_{ext} ~, \label{FQHEaction}
\end{equation}
where $F_{ext}=dA_{ext}$ is the external electromagnetic field
applied to the FQHE sample, while $A$ is the internal gauge field that
describes the low energy degrees of freedom of FQHE.

In a FQHE system, the parity symmetry is broken. On the other hand,
the original ABJM is parity-even since the two $U(N)$ gauge groups
are interchanged during a parity transformation. Therefore, we need
to break the parity symmetry of the ABJM theory in order to use it
to model FQHE. One way to achieve this is by adding $M$ fractional
D2-branes (i.e. $D4$-branes wrapped on $\mathbb{CP}^1$).  On the
gravity side, these $M$ fractional D2-branes are unstable and would
fall into the horizon of $AdS_4$, leaving only NSNS 2-form flux
behind \cite{ABJ}
\begin{equation}\int_{\mathbb{CP}^1}B_2=(2\pi)^2\f{M}{k}
~. \label{expb}
\end{equation}
On the field theory side, the gauge group $U(N)_k\times U(N)_{-k}$
changes into $U(N+M)_{k}\times U(N)_{-k}$, thus breaking the parity
symmetry. Treating the $U(N)_k\times U(N)_{-k}$ part which is common
to both sides as spectators and extracting the $U(1)\subset U(M)$
part of the Chern-Simons gauge theory, we arrive at $U(1)_k$
Chern-Simons action (first term in (\ref{FQHEaction})) that encodes
the low energy description of FQHE.

Adding D4-brane wrapped on $\mathbb{CP}^1$ breaks the parity
symmetry by shifting the rank of one of the two gauge groups.
Another way to break parity symmetry is to add $l$ D8-branes wrapped
on $\mathbb{CP}^3$. As shown in \cite{FWRT,GT}, it shifts the level
of one of the gauge groups: $U(N)_k\times U(N)_{-k}$ changes into
$U(N)_{k+l}\times U(N)_{-k}$. However, we will not discuss this
system in the present work, leaving its application as a future problem.

In one of the models constructed in the recent paper \cite{FWRT},
FQHE was realized by inserting defect D4 or D8-branes which are
interpreted as edge states of the Chern-Simons gauge theory. Below
we will show that we can also model FQHE without adding edges
states, expressing everything purely in terms of RR-fluxes in the
bulk $AdS_4$. Moreover, we will find that an addition of D6-branes
enables us to describe a class of FQH plateau transitions.

\subsection{Background RR-Field as External Field}

The fractional D2-brane (namely D4-brane wrapped on $\mathbb{CP}^1$)
has the world-volume action
\begin{equation}
S_{D4}=-T_{4}\int d^5\sigma e^{-\phi}\s{-\det(g+2\pi
F)}+2\pi^2T_4\int C_1 \we F\we F ~,
\end{equation}
where $T_4=(2\pi)^{-4}$ in the unit $\al=1$ and $C_1$ is sourced by
the $k$ units of the D6-brane flux,
which leads to
$\int_{\mathbb{CP}^1} F_2 =2\pi k$. Integrating over the internal
$\mathbb{CP}^1$, the Chern-Simons term of the D4-brane becomes
\begin{equation}
S^{CS}_{D4} =\f{k}{4\pi}\int_{R^{1,2}} A\we dA ~.
\end{equation}
Thus we obtained the first term in the Chern-Simons action
(\ref{FQHEaction}) with the internal gauge field $A$. In other words,
this Chern-Simons theory is the $U(M)$ part of the $U(N+M)\times U(N)$
gauge groups.

To couple the internal gauge field $A$ to an external source
$A_{ext}$, i.e. to realize the second term in action
(\ref{FQHEaction}), we need to turn on some background RR-flux.
Recall that the original ABJM theory has background RR-flux \be
F_2=\f{2k^2}{R^3}\omega ~,\ \ \   \ti{F}_4(\equiv F_4-A_1\we
H_3)=-\f{3}{8}R^3 \ep_{AdS_4} ~,\ \ \ H_3=0  \ee in $AdS_4\times
\mathbb{CP}^3$. We need to modify the RR-flux such that it provides
an external gauge field that couple to $A$ living on fractional
D2-branes.

It is easy to see that we can simply turn on additional RR 3-form
potential of the form
\begin{equation}\label{extraRR3}
C_3=\f{4\pi k}{R^3}A_{ext} \we \omega ~, \end{equation} where the
1-form $A_{ext}$ lies inside $AdS_4$ and will serve as the external
gauge field.
This extra $C_3$ sources\footnote{ Note that the volume of
$\mathbb{CP}^n$ is given by
$\int_{\mathbb{CP}^n}\f{\omega^n}{n!}=\f{\pi^n}{n!}(2L)^{2n}$ in our
convention. } an additional Chern-Simons term on the fractional
D2-brane:
\begin{equation} \f{1}{(2\pi)^4}\int_{R^{1,2}\times
\mathbb{CP}^1} 2\pi F\we C_3=\f{1}{2\pi}\int_{R^{1,2}} A_{ext}\we F
~,
\end{equation}
which gives the correct coupling to the external field (i.e. the
second term of (\ref{FQHEaction})). Then after taking into account
the kinetic term, we successfully realize the FQHE system coupled to
the background RR-field $A_{ext}$ as defined by the action
(\ref{FQHEaction}).\footnote{Similarly, we can describe a FQHE
system on the D2-branes instead of on fractional D2-branes by
turning on some extra RR 1-form $C_1\equiv A_{D2}$. The 1-from
serves as the external gauge field, and it couples to the D2-brane
in the standard way: $\f{1}{2\pi}\int_{R^{1,2}} A_{D2}\we F$. In
this paper we will only discuss FQHE system living on the fractional
D2-branes.}

A D-brane configuration modeling the FQHE has already been given in
\cite{BOB}, which is constructed from D0, D2, D6 and D8-branes (see
also \cite{Suss} for other brane model of FQHE). This model looks
similar to ours in the sense that a RR field plays the role of the
external magnetic field in FQHE.\footnote{We would like to thank O.
Bergman for pointing this out to us and for discussing possible
relations.} It would be very interesting to pursuit the relations
further as this may lead to the holographic construction of
\cite{BOB}.

\subsection{Holographic Dual}

Now we analyze the IIA supergravity with the modified RR-flux
profile.
We assume the following ansatz
\begin{align}
 F_2=\f{2k^2}{R^3}\omega + F_{D2} ~,\qquad
 \ti{F}_4=-\f{3}{8}R^3\ep_{AdS_4}+\f{4\pi k}{R^3}F_{ext}\we \omega ~,
\end{align}
and we require that the 3-form $H_{3}$ has indices only in the
$AdS_4$ directions. Moreover, $F_{D2}$ and $F_{ext}$ have indices in
the $AdS_4$ directions as well. We are interested in a combination
of these fluxes which become a massless gauge field \cite{ABJM}. In
this ansatz we find
\begin{align}
 *F_2=\f{R^3}{16}\ep_{AdS_4}\we\omega^2+*_4F_{D2}\we \f{\omega^3}{6} ~,\qquad
 *\ti{F}_4=\f{k^2}{R^3}\omega^3 +\f{2\pi k}{R^3}*_4F_{ext}\we \omega^2 ~,
\end{align}
where $*$ and $*_4$ denote the Hodge duals in the total
ten-dimensional spacetime and the $AdS_4$ spacetime, respectively.
The equations of motion of fluxes are written at the linearized
level as follows
\begin{align}
 dF_{D2}=0 ~,\qquad dH_{3}=0 ~, \qquad
 dF_{ext}=-\f{k}{2\pi}H_3 ~, \qquad
 d*_4F_{D2}=\f{6k^2}{R^3}H_3 ~, \no
 d*_4F_{ext}=0 ~,\qquad
 \f{1}{g_s^2}d*_4H_3=-\f{24\pi k^3}{R^6}*_4F_{ext}-\f{6k^2}{R^3}F_{D2} ~. \label{eomr}
\end{align}
It is easy to see that the mode
\be
F_{D2}=-\f{4\pi k}{R^3}*_4F_{ext} ~,\ \ \ H_3=0 ~, \label{masslessm}
\ee
becomes a massless 2-form field strength. Under this constraint,
the equations of motion (\ref{eomr}) become exact even beyond the linear order approximation.

Now we assume that the background includes $M$ fractional D2-branes,
which correspond to the NSNS 2-form (\ref{expb}). If we concentrate
on the massless mode (\ref{masslessm}), we find that the type IIA
action is reduced to \be S_{ext}=-\f{R^3}{48\pi^2 k}\int_{AdS_4}
F_{ext}\we
*F_{ext}-\f{M}{4\pi k}\int_{AdS_4} F_{ext}\we F_{ext}
~,\label{gauge} \ee where the second term comes from the
Chern-Simons term of the IIA supergravity $-\f{1}{4\kappa^2}\int
B_2\we dC_3\we dC_3.$
In the second term of the action, the topological term $\int
F_{ext}\we F_{ext}$ leads to a boundary Chern-Simons term $\int
A_{ext}\we F_{ext}$ in the AdS/CFT procedure (see also \cite{KKr}).
Since $A_{ext}$ is the external gauge field probing FQHE, we can
immediately read off the fractionally quantized Hall conductivity:
\be \sigma_{xy}=\f{M}{2\pi k}=\frac{M}{k}\cdot\frac{e^2}{h} ~, \ee
where we have restored the electron charge $e$ and $\hbar=1$.

In this way we have shown that the ABJM theory with $M>0$ fractional
D2-branes can model fractional quantum Hall effect. Something
interesting also happens at $M=0$. If we focus on the gauge theory
part while ignoring the gravity part, the theory as given by action
(\ref{gauge}) with $M=0$ has an S-duality that inverts the
Yang-Mills coupling,  as was also noted in a different example
\cite{Son}. Since the coupling is given by
$g_{YM}=\sqrt{\frac{12\pi^2k}{R^3}}\sim (\frac{k}{N})^{1/4}$, this
``S-duality'' exchanges the level $k$ and the rank $N$ in the ABJM
theory.

\subsection{Flavor D6-branes and Quantum Hall Transition}

In the previous system the Hall conductivity is fractionally
quantized, and the system describes one plateau of FQHE. In order to
describe the plateau transition in FQHE, where $\sigma_{xy}$ changes
continuously, we would like to add flavor D6-branes in this setup
and consider its deformation. Though the mechanism in our transition
described below is very similar to the one in D3-D7 systems
discussed in \cite{DKS}, its interpretation is different. This is
because in our case we regard the RR field $A_{ext}$ as the external
gauge field, whereas in \cite{DKS} the external gauge field is given
by the gauge field on the D7-brane. This is the main reason why we
can realize the plateau-transition for the fractional QHE, while the
paper \cite{DKS} realized the transition in the integer QHE.

In the calculation of $\sigma_{xy}$, new contributions essentially
come from the Chern-Simons terms of the D6-branes. We consider a
D6-branes wrapped on $S^3/{\mathbb Z}_2$ (\ref{thsp}) in the
presence of NSNS $B$-field (\ref{expb}). Its non-trivial
Chern-Simons terms are
\begin{equation}
S_{D6-RR}=\f{1}{(2\pi)^5}\int C_3\we F\we
B_{NS}+\f{1}{2(2\pi)^4}\int C_1\we B_{NS}\we F\we F ~.
\end{equation} Plugging in the explicit forms of $B_{NS}$ and $C_3$, we
find
\begin{equation}
S_{D6-RR}=\f{M}{2\pi k}\zeta \int_{R^{1,2}} A_{ext}\we F
+\f{M}{4\pi}\zeta \int_{R^{1,2}} A\we F ~,
\end{equation}
where $A$ is the $U(1)$ gauge field on the D6-brane and $A_{ext}$ is
the external field induced by the RR 3-from potential. The quantity
$\zeta$ is defined by
\begin{equation} \int_{[D6]} \omega \we \omega
=16\pi^2 L^4\zeta ~,
\end{equation} where $[D6]$ is the
four-dimensional worldvolume of the D6-branes in the $\mathbb{CP}^3$
directions. Here we normalized such that $\zeta=1/2$ when the
D6-brane wraps the following 4-cycle
\begin{equation} 0\leq
\xi<\f{\pi}{4} ~, \ \ 0\leq  \psi<2\pi ~,   \ \ 0\leq \theta<\pi ~,\
\ \ 0\leq \vp<2\pi ~. \end{equation} After combining $S_{D6-RR}$
with the boundary Chern-Simons term coming from the topological term
in (\ref{gauge}) and classically integrating out $A$, we finally
obtain
\begin{equation}
S_{tot-RR}=\f{M}{4\pi}\left(\f{1}{k}-\f{\zeta}{k^2}\right)
\int_{R^{1,2}} A_{ext}\we F_{ext} ~. \label{halle}
\end{equation}

Now we put the system at the finite temperature. The $AdS_4$ is then
replaced by a black hole solution. We can find (at least
numerically) solutions whose embedding function $\xi(r)$ changes
smoothly from $\xi(\infty)=\f{\pi}{4}$ to $\xi(r_0)=0$ for certain
large enough $r_0$. Notice that the point $\xi=0$ is interpreted in
the IIB brane configuration as where the flavor D5-brane and the
$(1,k)$5-brane make a ${\cal N}=3$ supersymmetric bound-state i.e.
the $(1,k+1)$5-brane as is seen from (\ref{changea}). A large value
of $r_0$ corresponds to large mass of the hypermultiplets. For
flavor masses large enough, the D6-brane does not touch the horizon.
As we reduce $r_0$ (or equivalently the flavor mass), the D6-brane
will move closer to the horizon and only stay away from it above
some critical value of $r_0$. Then if we reduce the flavor mass
further, the D6-brane will terminate at the horizon as is known in
the D3-D7 system \cite{MMT}. When this happens, the value of $\zeta$
jumps from $\zeta=1/2$ (for the D6-brane separated from the horizon)
to a certain value $\zeta=\zeta_0<1/2$, as in the D3-D7 system of
\cite{DKS}. As the flavor mass becomes smaller, $\zeta$ gets smaller
and finally reaches $\zeta=0$, which corresponds to the original
flavor D6-brane. This describes half of the transition process, and
the other half can be found similarly. Therefore, as we change the
flavor mass, the dual FQHE system undergoes a plateau transition
from $\nu=\f{1}{k}$ to $\nu=\f{1}{k+1}$ (recall that we assumed that
$k$ is large in (\ref{halle}) so that the description of type IIA
superstring theory is reliable). If we combine two D6-branes, then
we can realize more realistic transitions from $\nu=\f{1}{k}$ to
$\nu=\f{1}{k+2}$.\footnote{The elementary particles in realistic
models are either purely fermions as in traditional 2DEG FQHE
systems or purely bosons as in the more recent bosonic FQHE in
rotating cold atoms. Thus during a plateau transition $k$ jumps only
among odd integers if in fermionic systems, and only among even
integers if in bosonic ones.} It will be an interesting future
problem to examine the above transition in more detail and
calculate how $\sigma_{xx}$ and $\sigma_{xy}$ change
explicitly.

Finally notice that in the CFT side, this shift of the level can be
understood as the parity anomaly by adding mass to the
hypermultiplets and integrating them out. As mentioned in \cite{Ber}
there are three supersymmetric and one non-supersymmetric mass
deformations. The former correspond to the shifts of $\vec{x}_2$ in
the IIB brane configuration and appear in the fluctuation spectrum
of scalar modes in section 3. The parity anomaly only occurs in the
latter one. Therefore we expect our D6-brane configuration assumed
in this subsection to be non-supersymmetric.

\section{Conclusion}

In this paper, we have performed an analysis on probe branes dual to
the flavors in the ${\cal N}=6$ Chern-Simons theory, i.e., the ABJM
theory. We found that the probe branes are wrapped over $AdS_4
\times S^3/{\mathbb Z}_2$ in the dual geometry of $AdS_4 \times
\mathbb{CP}^3$. These are classified into two types by the ${\mathbb
Z}_2$ Wilson line and each corresponds to a flavor for each of the
two $U(N)$ gauge groups. The probe D6-brane is shown to preserve 12
supersymmetries, which are the same supersymmeties of the dual
${\cal N}=3$ superconformal symmetry. The brane configuration is
also confirmed by the analysis of a type IIB brane configuration
dual to the ABJM theory with flavors. We obtained the spectrum of
BPS mesonic operators in the ABJM theory with flavors by analyzing
the fluctuations of the dual D6-branes and found agreements with our
expectation. We also considered an application of the ${\cal N}=6$
Chern-Simons theory to the fractional quantum Hall effect. In the
presence of fractional D2-branes, we showed that it offers us a
simple holographic setup of fractional quantum Hall effect.
Moreover, we found that mass deformations of the flavor D6-brane are
interpreted as plateau transitions of fractional quantum Hall
effect.

There are several future directions we would like to consider. First
of all, we have only analyzed the flavor branes preserving the
maximal supersymmetry, and it would be important to look for branes
preserving less supersymmetry. Moreover, there are other
supersymmetric Chern-Simons theories with holographic duals such as
the one with the orientifold discussed in \cite{ABJ}. Therefore it
is possible to extend our analysis of flavor D-branes to these
cases. For the purpose of application to condensed matter physics,
it is very interesting to explicitly compute the conductivities in
our setup at finite temperatures. Furthermore, the relation to the
description of the plateau transition using the Chern-Simons
Ginzburg-Landau model \cite{KLZ} should also be clarified.

\vskip5mm

\noindent {\bf{Note added:}}
While we were preparing the draft, we noticed that
the paper \cite{Ingo} appeared in the arXiv.
It has a major overlap on the discussions of D6-branes
as the holographic dual of adding
flavors in the ${\cal N}=6$ Chern-Simons theory, though the identification is
slightly different. 
 After our paper appeared
on the arXiv, we found the paper \cite{GJ} listed on the same day, which also 
 argued the same interpretation on the flavor D6-branes as ours.

\vskip5mm

\noindent {\bf Acknowledgments}

We would like to thank T. Nishioka, Y. Okawa and S. Ryu for useful discussions.
We are also very grateful to Oren Bergman for valuable comments on
this paper. The work of YH is supported by JSPS Research Fellowship.
The work of WL and TT is supported
by World Premier International Research Center Initiative (WPI Initiative), MEXT, Japan.
The work of TT is also supported by JSPS Grant-in-Aid for Scientific Research No.20740132, and
by JSPS Grant-in-Aid for Creative Scientific Research No. 19GS0219.

\vskip2mm

\appendix

\section{Supersymmetry and Brane Configuration}
\label{susy}

Here we count the number of supersymmetries preserved by the type
IIB brane configuration system discussed in section \ref{Brane}. It
is convenient to combine the two chiral spinors (16 components each)
$\ep_L$ and $\ep_R$ in the type IIB supergravity by a complex 16
component spinor as $\ep=\ep_L+i\ep_R$. In this combination it
satisfies the chirality condition as \be \gamma_{0123456789}\ep=\ep
~.  \label{chiral} \ee Let us study which kind of supersymmetry is
preserved in the presence of branes. For a D3-brane in the
$0126$-directions, preserved supersymmetry is associated with the
spinor satisfying \be \ep=-i\gamma_{0126}\ep ~. \label{susyone} \ee
Similarly we have the constraint \be \ep=i\gamma_{012345}\bar{\ep}
~,  \label{susytwo} \ee for a D5-brane in the $012345$-directions
and \be \ep=\gamma_{012789}\bar{\ep} ~.  \label{susythree} \ee for a
NS5-brane in the $012789$-directions.

First we show that this D3-D5-NS5 system preserves
$1/4$ of the 32 supersymmetries.
Indeed, we can derive one of the three conditions (\ref{susyone}),
(\ref{susytwo}) and (\ref{susythree}) from the other two.
Diagonalizing the spinor by the actions of $\gamma_{37}$, $\gamma_{48}$ and
$\gamma_{59}$ as
\be
\gamma_{37}\ep=is_1\ep ~,  \ \ \ \ \gamma_{48}\ep=is_2\ep ~,\ \ \ \  \gamma_{59}\ep=is_3\ep ~,
\ee
the chirality constraint (\ref{chiral})  leads to
\be
\gamma_{0126}\ep=i(s_1s_2s_3)\ep ~,\ \ \ \ \gamma_{0126}\bar{\ep}=-i(s_1s_2s_3)\bar{\ep} ~.\label{susyfour}
\ee
The degrees of freedom of spinor is specified by $(s_1,s_2,s_3)$ and this leads to
8 complex components. Moreover, the condition (\ref{susyone}) requires
\be
s_1s_2s_3=1 ~,  \label{susyfive}
\ee and thus we have the following 4 possibilities
\be
(s_1,s_2,s_3)=(+++),(--+),(-+-),(+--) ~. \label{singc}
\ee
In this way the D3-D5-NS5 system preserves $1/4$ of the 32 supersymmetries due to the constraints
(\ref{susythree}) and (\ref{susyfive}).

In order to construct the brane configuration in section
\ref{Brane}, we further insert a $(1,k)$5-brane. We assumed that it
is rotated by the same angle $\theta$ in each of $37$, $48$ and $59$
planes. If we choose that this angle $\theta$ is given by
$\sin\theta=k/\s{1+k^2}$ (assuming $g_s=1$ and vanishing axion for
simplicity), then we find the following supersymmetry constraint on
the spinor as \be \ep=e^{i\theta}\cdot \gamma_{012789}\cdot
e^{-\theta(\gamma_{37}+\gamma_{48}+\gamma_{59})}~\bar{\ep} ~.
\label{susysix} \ee Suppose that $\ep_{(0)}$ satisfies the
supersymmetric condition for the D3-D5-NS5 system. Then, with the
help of (\ref{susythree}), we can find that the following spinor
$\ep$ satisfies the condition (\ref{susysix}) as \be
\ep=e^{i\f{\theta}{2}+\f{\theta}{2}(\gamma_{37}+\gamma_{48}+\gamma_{59})}\ep_{(0)}
~. \ee Since the $\ep_{(0)}$ are the spinors for the original
system, we would like to find spinors that satisfies $\epsilon =
\epsilon_{(0)}$. They correspond to supersymmetries surviving after
adding a $(1,k)$5-brane. This is given by the choice
$(--+),(-+-),(+--)$ in the (\ref{singc}). Thus we have found that
the $(1,k)$5-brane further breaks $1/4$ out of the original 8
supersymmetries of D3-D5-NS5 system. In this way, we have shown that
this final system preserves 6 supersymmetries corresponding to
${\cal N}=3$ Chern-Simons theory with flavors.

\section{The Orbifold Model on $S^3/{\mathbb Z}_2$}
\label{orbifold}

In this appendix we analyze which modes of spherical harmonics
survive under the ${\mathbb Z}_2$ orbifold projection.
The symmetry of $S^3$ is $SO(4) \sim SU(2)_R \times SU(2)_L$
and the function can be labeled by the representation of
$SU(2)_R \times SU(2)_L$. We denote $(m , \bar m)$ as the
eigenfunction of $J^3_R,J^3_L$. For the scalar function
in the $(\frac{l}{2},\frac{l}{2})$ representation,
the both run the same range as
$m,\bar m = - l/2, - l/2 + 1, \cdots , l/2$.
The scalar function satisfies
\begin{align}
 D_i D^i Y^l (S^3) = - l (l+2) Y^l (S^3)~,
\end{align}
where $D_i$ are the covariant derivatives on $S^3$.
In our case, the identification is taken for the shift
$\psi \to \psi + 4\pi/p$ with $p=2$, and the corresponding
orbifold model is obtained by the projection operator
(see, e.g., \cite{LM,Hikida})
\begin{align}
 P = \tfrac12 (1 + e^{4 \pi i  J^3_L /2} ) ~.
 \label{projection}
\end{align}
Therefore, only $2 \bar m \in 2 {\mathbb Z}$ survives under
the orbifold projection, which implies $l \in 2 {\mathbb Z}$
for the scalar function.  The vector spherical harmonics
are in the
$(\frac{l\pm 1}{2},\frac{l\mp 1}{2})$ representation,
and satisfy
\begin{align}
 D_i D^i Y^{l,\pm}_j (S^3) -R^k_{ ~j} Y^{l,\pm}_k (S^3)
   = - (l+1)^2 Y^{l,\pm}_j (S^3)
\end{align}
with  $R^k_{ ~j}$ as the Ricci tensor of $S^3$ .
The projection is the same as in the scalar case,
thus the restriction is $2 \bar m \in 2 {\mathbb Z}$,
which implies $l \in 2 {\mathbb Z} + 1$.

Let us consider the effects of ${\mathbb Z}_2$ Wilson loop
along the non-trivial cycle. We prepare two fractional branes
with and without ${\mathbb Z}_2$ Wilson loop.
The gauge symmetry is now $U(1) \times U(1)$,  and
there are two types of open string between
the same brane and between different branes.
The fields coming from the former do no feel any effect
of Wilson loop and the projection is the same as before.
For the other fields coming from the latter, the projection becomes
\begin{align}
 P = \tfrac12 (1 - e^{4 \pi i  J^3_L /2} )
\end{align}
due to the existence of Wilson loop. For this type of scalar field
the restriction is $2 \bar m \in 2 {\mathbb Z} + 1$, which exists
for $l \in 2 {\mathbb Z} + 1$. For this type of vector
field the restriction leads to $l \in 2 {\mathbb Z}$.


\end{document}